# Gas flow–directed growth of aligned carbon nanotubes from nonmetallic seeds

Yuanjia Liu [a], Taiki Inoue [a], Mengyue Wang [a], Michiharu Arifuku [b],

Noriko Kiyoyanagi [b], Yoshihiro Kobayashi [*a]

a Department of Applied Physics, Osaka University, Suita, Osaka 565-0871, Japan

b Nippon Kayaku Co., Ltd., Kita-ku, Tokyo 115-8588, Japan

*Email: kobayashi@ap.eng.osaka-u.ac.jp

Abstract

 Kite growth is a process that utilizes laminar gas flow in chemical vapor deposition to grow long, well-aligned carbon nanotubes (CNTs) for electronic application. This process uses metal nanoparticles (NPs) as catalytic seeds for CNT growth. However, these NPs remain as impurities in the grown CNT. In this study, nanodiamonds (NDs) with negligible catalytic activity were utilized as nonmetallic seeds instead of metal catalysts because they are stable at high temperatures and facilitate the growth of low-defect CNTs without residual metal impurities. Results demonstrate the successful growth of over 100-µm-long CNTs by carefully controlling the growth conditions. Importantly, we developed an analysis method that utilizes secondary electron (SE) yield to distinguish whether or not CNTs grown from metal impurities. The absence of metallic NPs at the CNT tips was revealed by the SE yield mapping, whereas the presence of some kind of NPs at the same locations was confirmed by atomic force microscopy (AFM). These results suggest that most of the aligned CNTs were grown from nonmetallic seeds, most likely ND-derived NPs, via the tip-growth mode. Structural characterizations revealed the high crystallinity of CNTs, with relatively small diameters. This study presents the first successful use of nonmetallic seeds for kite growth and provides a convincing alternative for starting materials to prepare long, aligned CNTs without metal impurities. The findings of this study pave the way for more convenient fabrication of aligned CNT-based devices, potentially simplifying the production process by avoiding the need for the removal of metal impurities.

# 1. Introduction

Carbon nanotubes (CNTs) [1,2] have garnered attention due to their excellent properties [3,4] and various promising applications [5–9]. In particular, aligned CNTs grown on surfaces by chemical vapor deposition (CVD) with macroscale lengths, low defects, and parallel alignment are ideal candidates for next-generation electronics applications [10,11] and have attracted research interest [12,13]. In the past decades, significant progress has been made toward the controlled synthesis of aligned CNTs on substrate surfaces [14–16]. Gas flow–directed growth, also known as kite growth, is a well-established method [17] and has several advantages over other methods, including ultralong CNT length and wide applicability of the as-formed air-suspended CNTs. The "kite mechanism" was first proposed by Huang et al. to explain the growth of aligned CNTs [18]. In this mechanism, convective effects caused by vertical temperature gradients overcome the catalyst–substrate interaction and lift both the catalyst at the CNT tip and the CNT; then, the horizontal laminar flow carries the CNTs and aligns the growing CNTs [19]. During the growth process, the catalyst floats above the substrate and moves with the growing CNT tip, also known as the tip-growth mode. In contrast, in base growth, the catalyst is at the root of the CNT and does not move with the growing CNT. During tip growth along with kite growth, the strong van der Waals interaction between CNTs and substrates fixes the originally floating CNTs on the substrates after their growth. Significant advances have been made recently toward the controlled synthesis of CNTs by kite growth due to continuous research efforts [20–23].

Catalyst nanoparticle (NP) preparation, particularly the selection of their elements, has a significant impact on the growth behavior of CNTs via CVD. A typical kite-growth process uses metal NPs as catalysts, such as Fe [18], Co [24], and Cu [25]. Typical transition metal catalysts, such as Fe and Co, mediate the CNT growth through their NPs in the liquid phase [26]. CNT growth from liquid NPs usually occurs via a so-called vapor–liquid–solid (VLS) mechanism [27], where carbon atoms are supplied to the CNT through the bulk diffusion of carbon in the NPs. These molten NPs tend to deform during the synthesis of CNTs at high temperatures, which is considered the main origin of defect formation in CNTs [28]. The CNT tips connected to catalysts may not be used for usual device fabrication. However, during growth, metal catalysts may be once evaporated in the chamber and redeposited on CNTs, resulting in contamination by metal NPs as reported in a literature [23,29]. Purification processes are necessary to remove these contaminants before further utilization, but such processes often introduce additional defects in the CNTs[30]. A possible solution to overcome the drawbacks of metal catalysts is to grow CNTs from nonmetallic NPs. Hereafter, the NP used for CNT growth is referred to as the "growth seed" instead of a "catalyst"; the nonmetallic NP has negligible catalytic activity but plays a role as a template for CNT growth [31]. As a representative nonmetallic growth seed, nanodiamond (ND) [32–34] has attracted research interest due to its unique properties. As the diffusion of carbon in diamond is negligibly small compared to metals such as Fe [35,36], the bulk diffusion of carbon is unlikely to promote the growth of CNTs. Therefore, the growth of CNTs from ND must be driven by the surface diffusion of carbon, known as vapor–solid surface–solid (VSSS) mechanism [31,32]. ND as a growth seed offers excellent thermal stability at high temperatures and facilitates the growth of low-defect CNTs without any residual metal impurities. ND is also beneficial for device fabrication because it can facilitate CNT growth on any substrate type, including metal electrodes [32]. Despite these advantages, the kite growth of CNTs from ND has not been reported yet, presumably because of low CNT growth yields due to the limited catalytic activity of ND and narrow window of growth conditions. Moreover, CNT yield from the kite growth is low even when metal catalysts are used, and the co-influence of ND and kite growth further worsens this issue. To overcome this issue, we have attempted to further understand the CNT growth mechanism, especially the nucleation of cap structure at the initial growth stage [33,34]. This strategy can be extended to improve the yield of CNTs from ND via kite growth.

In this study, we achieved the kite-growth of aligned CNTs from nonmetallic growth seed, most likely ND-derived carbon NPs. After confirming the successful growth of aligned CNTs on ND-deposited substrates, we conducted a parametric study to investigate the effects of temperature and gas flow rate on the kite growth of CNTs. Scanning electron microscopy (SEM) is used to eliminate the possible contribution of metal impurities. Furthermore, the kite-growth mechanism is further evaluated using SEM and atomic force microscopy (AFM) images to determine whether the growth occurred via the tip-growth or base-growth modes. Finally, structural analysis of the aligned CNTs grown from nonmetallic NPs was performed using Raman spectroscopy and AFM, providing a comprehensive understanding of the structural properties of the synthesized CNTs.

## 2. Experimental section

### 2.1. Deposition of ND growth seeds

Purified ND particles prepared via the detonation method (original diameter: 4–14 nm) [37] and dispersed in ethanol (2.0 wt%) were used as the growth seeds to synthesize aligned CNTs. To support the ND particles, ~1 cm × 1 cm Si substrates with a 300-nm-thick thermal oxide layer were cleaned by ultrasonication sequentially with acetone, ethanol, and deionized water, followed by an ozone treatment process (L-UV253, Japan Electronics Industry) by flowing oxygen at a flow rate of 6 L/min under ultraviolet light for 48 min. After cleaning, adhesive tapes were applied to half of the substrate as a mask, and 10 μL of ND/ethanol solution was dropped on the other half of the substrate and dried naturally. After peeling off the tapes, followed by the heat treatment described in Section 2.2 to remove the glue, half of the substrate without ND was designated for aligned CNT growth.

### 2.2. Pretreatment of ND growth seeds

Herein, a 1040-mm-long quartz tube with a semicircular cross section of 44-mm diameter fixed in a tubular CVD furnace (GE-1000, GII Techno) with a heating length of 890 mm was used as the reactor. Figure 1 illustrates the growth process of long, aligned CNTs on the substrate via the kite-growth mechanism comprising two steps. In step 1, ND was pretreated for preparation as a growth seed. The quartz tube was heated to 600°C in advance without the samples; then, the substrates with ND particles were pushed into the quartz tube in the middle of the furnace and kept at 600°C in air for 10 min to remove any residual impurities. In this treatment, the diameter of the ND particles decreased to less than ~4 nm, which is the appropriate size for CNT growth [34].

### 2.3. Synthesis of aligned CNTs from ND growth seeds

Step 2 was the kite growth of CNTs, where optimized conditions were investigated by controlling the temperature and flow rate. After the heat treatment of the ND-deposited substrates (Fig. 1, step 1), the furnace temperature was increased to the synthesis temperature (850°C–950°C) under continuous 20-sccm $H_2$ (3%)/Ar flow at 85 kPa. The CNT growth was initiated by introducing 35 sccm $H_2$ (3%)/Ar bubbling through an ethanol reservoir at 40°C and 420 sccm $H_2$ (3%)/Ar through another gas line (Fig. 1, step 2). The growth process was conducted at atmospheric pressure, and the partial pressure of ethanol was estimated to be ~1.4 kPa. After 30 min of growth, the reaction was terminated by switching the gas to Ar. We employed relatively long growth time to compensate for the low efficiency of CNT growth from ND. Finally, the furnace was cooled to room temperature with a continuous flow of Ar. Fe catalyst was also used in the control experiment, as shown in Fig. S1.

2.4. Characterization

SEM (S-4800, Hitachi High-Tech) was used to characterize the grown CNTs (acceleration voltage of 1 kV and emission current of 10 μA, upper detector). The growth seeds/catalyst NPs at the tip of the grown CNTs were analyzed in detail based on the SE yield of SEM (acceleration voltage of 5 kV and emission current of 10 μA, upper detector); details are provided in Section 3.2, where we discuss how the SE yield allowed us to distinguish the elements located at the tip of the CNTs. The structure of the CNTs was characterized by Raman spectroscopy (RAMANtouch, Nanophoton) with an excitation laser wavelength of 532 nm and an objective lens of 100x and 0.90 NA, which results in a spatial resolution of approximately 350 nm. Dynamic force mode measurement of AFM (AFM5000, Hitachi High-Tech) was also used to characterize the grown CNTs and the growth seed/catalyst NPs at the tip of the CNTs.

To characterize grown CNTs at the same positions using multiple methods, such as SEM, AFM, and Raman spectroscopy, marker structures (Fig. S2) were patterned on the substrates after CNT growth using photolithography, subsequent metal deposition, and lift-off.

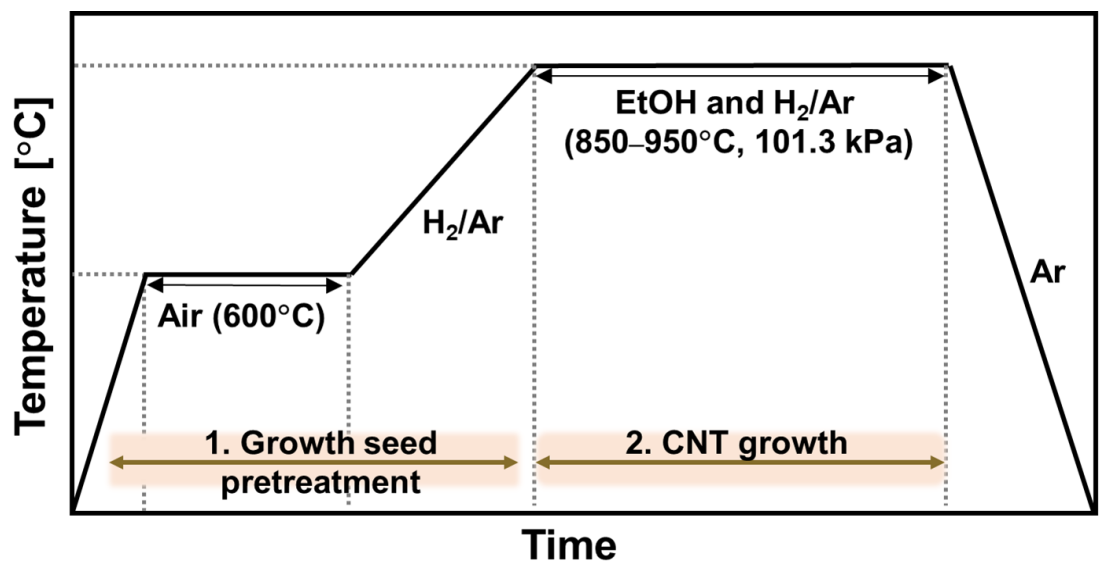

**Figure 1 Schematic of the pretreatment of ND growth seed in the reactor (step 1) and the subsequent kite growth of aligned CNTs on the substrates (step 2).**

## 3. Results and discussion

3.1 Temperature and gas flow rate dependence of CNT synthesis

We synthesized aligned CNTs on ND-deposited substrates using the aforementioned process under varying growth conditions. Figure 2 (a) shows the SEM images of aligned CNTs grown on $SiO_2$/Si substrates under optimized temperatures and gas flow rates. As shown, the aligned CNTs extended from the area of deposited ND particles and are longer than 100 μm. The alignment direction of the CNTs is parallel to that of the gas flow. These growth results are similar to those reported previously on kite growth of CNTs using metal catalysts [18,29,38–42]. Considering that the substrate is silicon with a thermal oxide layer, there is no possibility of crystal lattice–directed growth, indicating that the growth of the aligned CNTs follows the kite-growth mechanism. The confirmation of CNT growth from nonmetallic seed and exclusion of the contribution from metal impurities are provided in Section 3.2.

The occurrence of the kite-growth is further understood via fluid dynamics; therein, buoyancy lifts the CNT tips and then the laminar flow guides the stable lengthening of CNTs. Proper buoyancy and laminar flow are essential for stable kite growth. At the same time, CNT growth is governed by the activity of the growth seeds and the carbon supply rate [17], which are closely related to the growth rate even for the kite growth. Thus, the growth condition should be optimized comprehensively from these aspects. The buoyancy and laminar characteristics of the gas flow are evaluated by the Richardson number $Ri$ and Reynolds number $Re$. They can be expressed as $Ri = \Delta\rho g h/\rho v^2$ and $Re = \rho v d/\mu$, where $\Delta\rho, g, \rho, v, d,$ and $\mu$ represent the density difference in vertical length $h$, gravitational acceleration constant, gas density, gas flow velocity, inner diameter of the reactor tube, and coefficient of gas viscosity, respectively [43,44]. According to these formulae, as the process temperature $T$ increases, the $Ri$ increases while the $Re$ decreases. On the other hand, as the $v$ increases, the $Ri$ decreases while the $Re$ increases. These relationships are based on the equations of state assuming the ideal gas and the viscosity formula [45]. A decrease in the $Ri$ corresponds to a reduction in buoyancy, whereas a decrease in $Re$ promotes a more stable laminar flow. As per the aforementioned relation, achieving proper buoyancy requires an appropriate value of $Ri$ and minimum $Re$ to promote stable laminar flow. Two strategies can be employed to decrease the $Re$ for the efficient growth of aligned CNTs: reducing the gas flow velocity ($v$) and increasing the temperature ($T$). A larger buoyancy (large $Ri$) is simultaneously obtained; however, a too large buoyancy will cause strong gas convection. Therefore, optimizing the growth conditions is crucial to obtain a suitable buoyancy that can lift the growth seed and the grown CNTs into the laminar flow without affecting its stability. We thus examined the effects of the main factors, i.e., the temperature and gas flow rate, on the growth of aligned CNTs separately.

We investigated the dependence of temperature on the aligned growth of CNTs. CVD using ethanol as the carbon source is a widely adopted method for aligned CNT growth, owing to the etching effect of hydroxyl radical [46]. Initially, the temperature and gas flow rate were set to 950°C and 455 sccm, respectively, based on the process described by Liu et al. [41] However, in contrast to their study, we chose to employ a lower flow rate in order to achieve a more stable laminar flow under our experimental conditions. Aligned CNTs were grown at varying growth temperatures from 850°C to 950°C and a constant gas flow rate of 455 sccm to evaluate the average length and the yield of aligned CNTs. Two or more substrates with a size of approximately 1 cm × 1 cm were employed for each condition. We defined aligned CNTs as CNTs longer than 50 μm extending from the ND-deposited region that are grown in the gas flow direction. The yield of aligned CNTs was determined by calculating the average number of CNTs observed per substrate under a given condition. This was done by dividing the total number of aligned CNTs observed across all the substrates by the total number of substrates. SEM images of the aligned CNTs grown at different temperatures are shown in Fig. S3, and the average length and the yield of the aligned CNTs are plotted against temperatures in Fig. 2 (b). Considering fluid dynamics, higher temperatures cause more stabler laminar flows (small $Re$) but also possibly lead to excessive buoyancy (large $Ri$), whereas lower temperatures lead to more unstable flows (large $Re$). CNTs were not grown directionally below 900°C (Fig. S3 (a)). At 900°C, CNTs grew more than 200 μm in length but were not well aligned (Fig. S3 (b)). The CNTs grown between 930°C and 950°C were generally well aligned; the highest growth yield was obtained at 930°C (Fig. 2 (b)).

Gas flow rate dependence was examined at a fixed temperature of 930°C (Figs. 2 (c) and S4). The effect of gas flow rate on $Ri$ and $Re$ is similar to that of temperature. For a lower gas flow rate, $Re$ becomes small and the smaller $Re$ tends to cause stabler laminar flow. Contrarily, $Ri$ possibly becomes large, resulting in significant buoyancy, which is not suitable for stable kite growth. Higher gas flow rates lead to larger $Re$ values and cause more unstable flow. Thus, it becomes challenging to grow longer CNTs below 300 sccm (Fig. S4 (a)). Above 600 sccm, although long CNTs were grown in the direction of the gas flow, the degree of alignment was not high (Fig. S4 (e)). The optimal growth occurred within the range of 300 to 600 sccm, as the CNTs were longer and well aligned. The highest growth yield was realized at 455 sccm (Fig. 2 (c)).

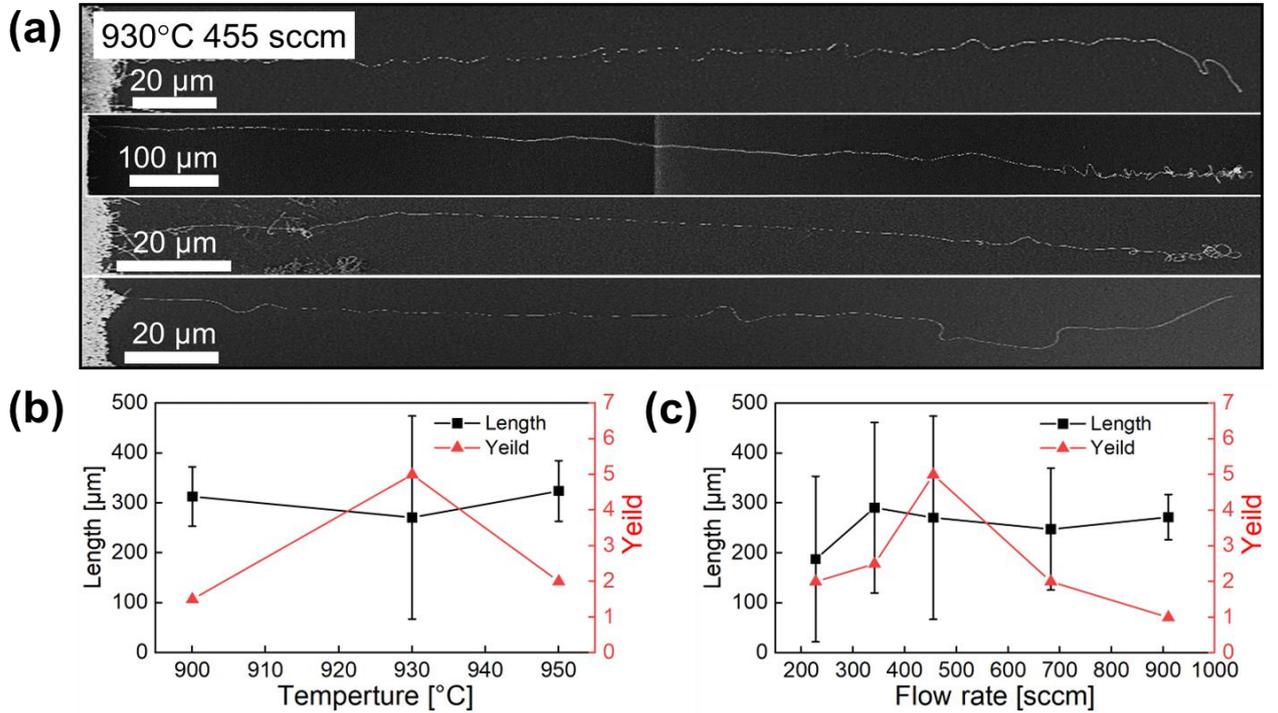

**Figure 2 (a)** SEM images of the aligned CNTs grown on ND-deposited substrates under the optimized condition (445 sccm and 930°C). The horizontal direction from left to right in the images corresponds to the direction of the gas flow. The ND-deposited areas look bright at the left end of the images. **(b, c)** The average length and the yield of aligned CNTs. CNTs are grown (b) at different temperatures with a flow rate of 455 sccm and (c) with different flow rates at 930°C. The yield was calculated as the average number of aligned CNTs observed per substrate (~1 cm × 1 cm size). The error bars represent the standard deviation of the length of all aligned CNTs measured for each condition.

3.2 Verification of nonmetallic particles as the origin of CNT growth

Since there remains the possibility that metal impurity in ND contributes to the CNT growth, the NPs at the tip of the aligned CNTs should be carefully examined. The ND solution used herein contained a small but not negligible amount of metal impurities due to the limitations of the production process, with Fe being one of the main components [33]; Fe is an active catalyst for CNT growth. Our previous study indicated that a small concentration of metal impurities (Fe, 80 ppm) in the ND samples do not contribute to the synthesis of CNTs at high temperatures [33,34]. In this study, a lower temperature was used compared to the previous studies, and thus, the impurity metal NPs may possess catalytic activity. Potential methods for identifying the element and structure of the growth seeds include transmission electron microscopy (TEM) and Raman spectroscopy.

However, TEM requires a suspended structure of the specimen, and preparing samples to observe the tips of long CNTs is highly challenging. Raman spectroscopy also presents difficulties when applied to individual NPs due to their minute size, which results in an extremely weak signal. Thus, it is difficult to directly distinguish the elements of the NPs at the tip of the CNTs on the substrates. For SEM observation, secondary electrons (SEs) are the most widely used probe in imaging due to their high sensitivity to surface topography and specimen composition [47]. As the elemental difference of carbon and Fe would result in different SE yields, we used SE yield mapping to detect the presence of impurity Fe NPs. Takagi et al. utilized SE yield to investigate the existence of metal catalysts on CNTs [48]. In this study, we extended this method to confirm the presence of nonmetallic seeds.

The SE yield of $SiO_2$ is higher than that of carbon at any given acceleration voltage [47]. However, a $SiO_2$ substrate usually appears darker than a CNT in SEM images. SEM observation of CNT is generally performed at a primary electron (PE) energy of around 1 keV. At this energy, the SE yield of $SiO_2$ is >1 (Fig. S5). Under continuous irradiation of the electron beam, the electrons emitted from the irradiated area as SEs are not supplemented by electrical conduction from surrounding unirradiated areas due to the insulating nature of $SiO_2$. This results in a positive charge-up on the irradiated area of the $SiO_2$ substrate, which hinders the emission of SEs with a negative charge from the $SiO_2$ surface and causes a darker contrast although the intrinsic SE yield of $SiO_2$ is relatively high. However, when the CNT is in direct contact with the $SiO_2$ substrate, electrons are supplied from the CNT to the substrate surrounding it through the electron beam-induced current (EBIC), resulting in the recovery of the SE yield. This leads to a brighter contrast of the substrate around the CNT, making the CNT appear thicker; this width depends on the EBIC range [49]. While EBIC imaging is useful for observing CNT, the intense SE emission of the substrate around the CNT can make it difficult to observe a NP connected to the CNT tip. Thus, the effect of the EBIC on the CNT should be eliminated. As the PE energy increases, the EBIC range also increases until it reaches the Si conductor substrate underneath the $SiO_2$ layer. At this point, electrons are supplied from the Si substrate to the entire $SiO_2$ surface through the EBIC, which recovers the SE yield and makes the entire substrate appear with the same contrast. To ensure the penetration of the 300-nm-thick $SiO_2$ layer and eliminate the effect of EBIC, a higher PE energy could be used. However, this may result in significant damage to the CNTs [50]. Therefore, a PE energy of 5 keV was selected as it reduces the effect of EBIC current while minimizing damage to the CNTs (Fig. S5). Since the SE yield of Fe is higher than that of carbon [47], it is easy to confirm the presence and absence of Fe NPs at CNT tips at a PE energy of 5 keV.

Both ND and CNT are composed of carbon, and although they have different SE yields due to their different structures, they have approximately the same SE yield [51]. Their SE yields are lower than that of Fe, thus resulting in CNT tips with a similar contrast to their bodies if CNT tips are connected with ND particles or free from any NPs, including metals.

Based on the discussion above, we investigated the presence of impurity metal NPs at the CNT tips by SE yield mapping at an accretion voltage of 5 kV (PE energy of 5 keV). Figure 3 (a) shows the SEM images and SE intensity profile of a typical CNT synthesized by depositing ND growth seeds. We also conducted the same observation of the control samples synthesized by depositing Fe catalysts (Fig. 3 (b)). Care was taken to avoid the saturation of the brightness at the tips of the CNTs during the SE yield map measurements. In the subsequent discussions, the terms "bright" and "dark" are used to describe CNT tips, which respectively imply that a CNT tip is observed to be brighter than and as dark as the corresponding CNT body in SE yield mapping. For a concise expression of the CNT samples, the term "ND-deposited samples" and "Fe-deposited samples" are used to refer to the samples grown from ND, which may contain metal impurities, and the samples grown from Fe, respectively. For the ND-deposited samples, dark tips were frequently observed from long CNTs (Fig. 3 (a)). For the Fe-deposited samples, bright particles were observed at the tips of long CNTs (Fig. 3(b)), whose appearance was consistent with that reported in a previous study on SE yield mapping of CNTs grown from metal catalysts [52]. The bright tips indicate that Fe NPs are attached to the CNT tips because the SE yield of Fe is higher than that of carbon. Figures 3 (c) and (d) summarize the observed ratio of bright and dark tips of CNTs for the samples grown from ND and Fe. We first plotted the ratio of bright and dark tips for all CNTs indicated as "total" in the figures. For the Fe-deposited samples, although many CNT tips were bright due to the Fe NPs at the tip, a small portion of the CNT tips was found to be dark ("total" in Fig. 3 (d)). We infer that these CNTs with dark tips are not grown by the tip-growth mode but by the base-growth mode because of their relatively short length. Observation of approximately 30 CNTs on the Fe-deposited samples revealed that all the CNTs with dark tips were <50 μm. Thus, we defined 50 μm as the boundary to distinguish long and short CNTs and separately plotted the ratio of bright and dark tips in Figs. 3 (c) and (d) for "short" and "long." In the case of Fe-deposited samples, both the tip- and base-growth modes coexisted for the short CNTs ("short" in Fig. 3 (d)), whereas only the tip-growth mode exists for the long CNTs ("long" in Fig. 3 (d)). In the case of the ND-deposited samples, most of the CNTs showed dark tips ("total" in Fig. 3 (c)). There are two possible explanations for the dark tips for the ND-deposited samples: CNTs grown via the tip/base-growth mode from nonmetallic NPs and CNTs grown via the base-growth mode from metal impurities. Because long CNTs over 50 μm with Fe catalyst were always grown via the tip-growth mode ("long" in Fig. 3 (d)), the possibility of long CNT growth via the base-growth mode from metal impurities was excluded. Thus, CNTs longer than 50 μm with dark tips were grown from nonmetallic growth seeds ("long" in Fig. 3 (c)). Some bright tips existed for ND-deposited samples regardless of the lengths, which should be attributed to CNTs grown via the tip-growth mode from metal impurities ("short" and "long" in Fig. 3 (c)).Therefore, for the ND-deposited samples, it is concluded that 80% of the long CNTs were grown not from metal impurities but from nonmetallic seeds via the kite-growth mechanism.

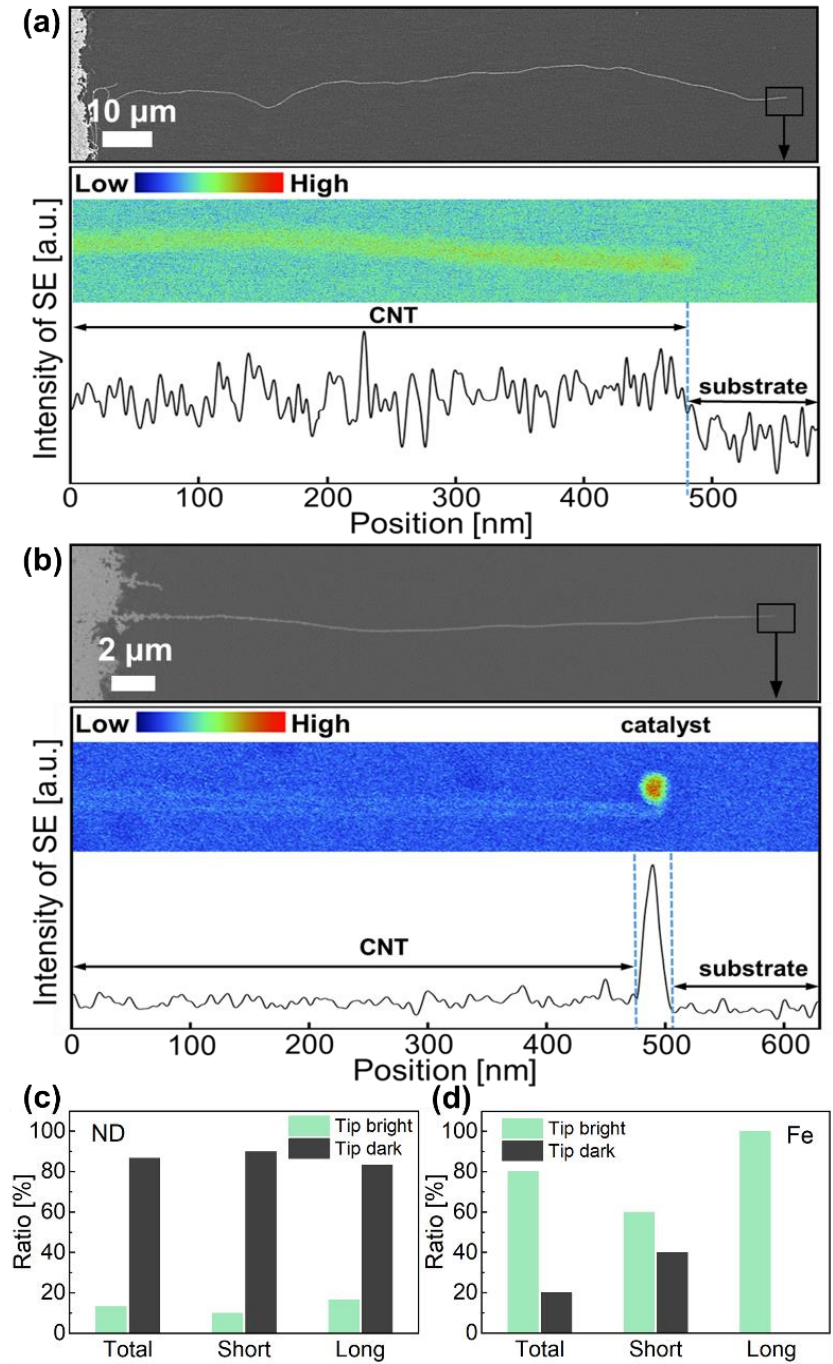

**Figure 3** SEM images of typical CNTs for (a) ND-deposited samples and (b) Fe-deposited samples measured at a PE energy of 1 keV. The images show the SE yield maps of the magnified CNT tips in color measured at a PE energy of 5 keV. Ratio of bright/dark CNT tips for (c) ND-deposited samples and (d) Fe-deposited samples with different CNT lengths. Total represents the sum of all CNTs. Short and long, respectively, denote CNTs with lengths shorter and longer than 50 μm.

Although we confirmed the absence of metallic NPs at the tip of aligned CNTs, it is challenging to directly confirm nonmetallic NP attachment using SEM because both ND-derived NPs and CNTs are composed of carbon. Thus, the presence of nonmetallic NPs is further verified. To clarify the growth mode of the CNTs grown from nonmetallic NPs (tip- or base-growth modes), AFM was employed to analyze topographic images around CNT tips on the nanometer scale in combination with SEM observation. Figures 4 (a) and (b) show the SE yield map and AFM image of a tip of an identical CNT longer than 50 µm (more information in Fig. S6). For the SE yield map, the tip of the CNT is dark, indicating the absence of metal NPs. For the AFM image of the same location on the CNT, a 5-nm-high protrusion is found at the CNT tip, indicating that the CNT was grown from nonmetallic NPs using the tip-growth mode. Due to the aforementioned challenges and the lack of direct observations such as TEM, a definitive characterization of the nonmetallic NPs at the tip of CNTs cannot be provided. Considering that other candidates of nonmetallic NPs, such as amorphous carbon, are unlikely to serve as a seed for CNT growth, it is believed that there is a strong possibility that the nanoparticles at the CNT tips are NDs or ND-derived NPs. By observing other CNTs, we elucidated that ~95% of the aligned CNTs longer than 50 µm grown from nonmetallic NPs follow the tip-growth mode. A small portion (~5%) of the CNTs exhibited dark tips in the SE yield maps, without distinct protrusions in AFM images. The most plausible hypothesis is that the diameters of the CNTs are comparable with those of the NPs [52], making it difficult to discriminate the particles from the CNT by AFM (Fig. S7). Another possibility is the detachment of NPs just before growth termination [53], which has been reported in the case of metal catalysts.

It is noteworthy that the diameters of CNTs grown via the kite growth are consistently smaller than the size of the nonmetallic NP, indicating that the maximum CNT diameter is limited by the size of the nonmetallic NP (Fig. 4 (c)). This finding is consistent with previous research on CNT growth using metal catalysts [48,54]. Furthermore, these results suggest that the CNTs in this study were grown not in the tangential mode but in the perpendicular mode [52]. The relationship between nonmetallic NP size and CNT diameter appears to be independent, which is consistent with previous study on solid catalyst [55].

Figure 5 shows the length distribution of grown CNTs from nonmetallic NP growth seeds with the different states of the tips. Type I represents the CNTs with tips showing dark contrast in SE yield maps and particle structures in AFM images, corresponding to the CNT grown from nonmetallic NPs via the tip-growth mode. Type II is the CNT with a dark SE contrast and no distinct protrusion structure at the tip, which is considered as tip-growth from nonmetallic NPs but with no distinct diameter difference between the particle and CNT. Type III is the CNT with bright SE contrast at the tip, which is regarded as tip growth from metal impurities. After distinguishing types II and III from type I, the lengths of CNTs grown from nonmetallic NPs via the tip-growth mode (type I) are found to range from 50 to 900 μm. In contrast, the CNTs grown from ND in previous studies [32–34] were just below 5 μm, highlighting the significant advantages of the kite-growth mechanism in the synthesis of longer CNTs. The distribution of CNT lengths in this study indicates a potential accordance with the Schulz-Flory distribution [17] (Fig. S8), but more sample numbers are needed for detailed discussion. The relationship between the diameter and length of CNTs has also been studied, but there is no direct correlation between them (Fig. S9).

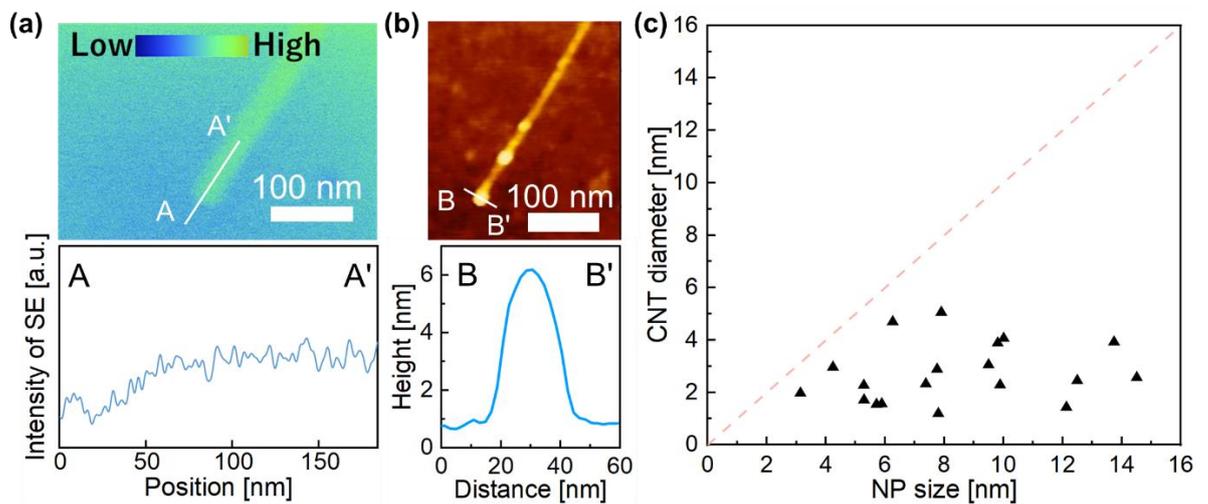

**Figure 4 (a) SE yield map and line profile of an aligned CNT tip. The line profile was obtained along a segment of the SE yield map represented by a line between two points, denoted as A and A′. (b) AFM image and line profile of the same CNT tip. The line profile was obtained along a segment of the AFM image represented by a line between two points, denoted as B and B′. (c) The distribution of nonmetallic NP sizes and the corresponding diameters of CNTs grown via kite growth.**

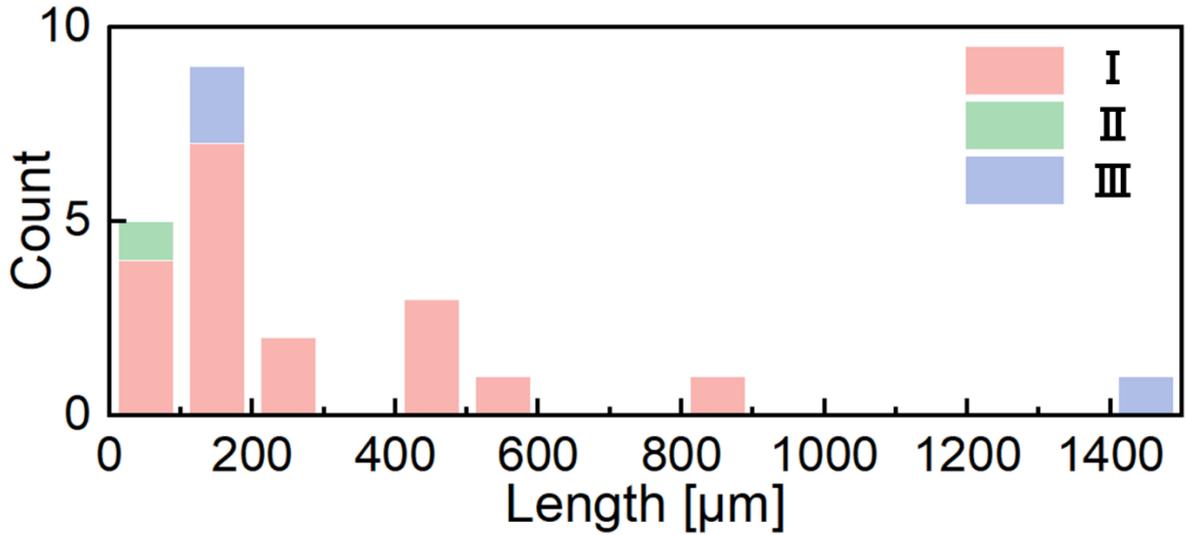

**Figure 5** Length distribution of CNTs formed by kite-growth mechanism on ND-deposited samples with different types of tips. Type I corresponds to CNTs with a dark SE contrast and distinct protrusion structures at the tips (tip growth from nonmetallic NPs). Type II indicates a CNT with a dark SE contrast and no distinct protrusion structures at the tip (presumably tip growth from nonmetallic NPs whose diameter is comparable with that of CNT). Type III indicates CNTs having tips of bright SE contrast (tip growth from metal impurities).

3.3 Crystallinity and diameter analysis of CNTs grown from nonmetallic NPs

We characterized the structure of aligned CNTs longer than 50 μm grown from nonmetallic NPs by SEM and Raman spectroscopy, as shown in Figs. 6 (a) and (b). Structural analysis was performed using the Raman mapping image of a long CNT (longer than 100 μm) in comparison with the SEM images observed at the same position. The G-band mapping image shows the same shape as the SEM image. The Raman spectra at the three positions indicated by the circles in the G-band image (Fig. 6 (b)) are shown in Figs. 6 (c) and (d). The spectra are consecutive but split into two images for clarity and accurate presentation. The intensity ratio from G-band to D-band in the Raman spectra (Fig. 6 (c)) was evaluated as 260 ± 50, indicating that the aligned CNTs uniformly have high crystallinity. The RBM peaks of the CNT were measured at three different positions, (i)–(iii), and it was found that all three positions displayed an identical peak at 172 cm$^{-1}$ (Fig. 6 (d)). If the diameter is calculated using the formula $\omega_{\text{RBM}} = A/d_t + B$, with $A = (217.8 \pm 0.3)$ cm$^{-1}$ nm and $B = (15.7 \pm 0.3)$ cm$^{-1}$ [56], the RBM signal at 172 cm$^{-1}$ corresponds to a diameter of 1.39 nm. The RBM frequencies are uniform in the entire imaging range; thus, the diameters of the carbon nanotubes are constant over 50 μm.

The diameter of the CNT shown in Figs. 6(a)–(d) was evaluated near its tip using AFM.

Figures 6 (e) and (f) show the AFM image and its line profile. The height of the CNT in the AFM image is 2.1 ± 0.1 nm, which is much larger than the height expected from the RBM signal. We have two possible explanations for this observation. The first involves a tight bundle of single-walled CNTs, in which only one CNT resonates with the excitation laser. The second possibility is that the two diameter values observed as CNT diameters represent the diameters for the outermost tube and one of the inner tubes in a multiwall CNT. Herein, the second possibility is considered because the bundle formation in the first possibility is rather unlikely due to the very low CNT yield from kite growth. Assuming that two shells with the diameters determined from AFM and RBM form a concentric structure, the interlayer distance between the shells is calculated as 0.355 ± 0.05 nm, which is a reasonable value compared to the interlayer distance of graphite (0.34 nm). Therefore, we infer this long CNT as a double-walled CNT with inner and outer wall diameters of ~1.4 and ~2.1 nm, respectively. Note that the RBM frequency for the outer wall should be observed around 119 ± 5 cm$^{-1}$ from the diameter of the AFM image and the calculation using the above equation. The corresponding RBM peak, however, was not observed in the measured Raman spectrum although the frequency was within the measurable range of the Raman system used herein. Results indicate that the outer wall does not resonate with the Raman excitation laser, which is consistent with previous research results [57]. Figure 6 (g) shows the diameter distribution of CNTs longer than 50 μm from nonmetallic NPs evaluated by AFM. Most of the nanotubes are regarded as single- and double-walled CNTs with small diameters, and some are multiwalled CNTs, which confirm the effectiveness of nonmetallic growth seeds for growing CNTs with a relatively small diameter, which are suitable for electronic applications.

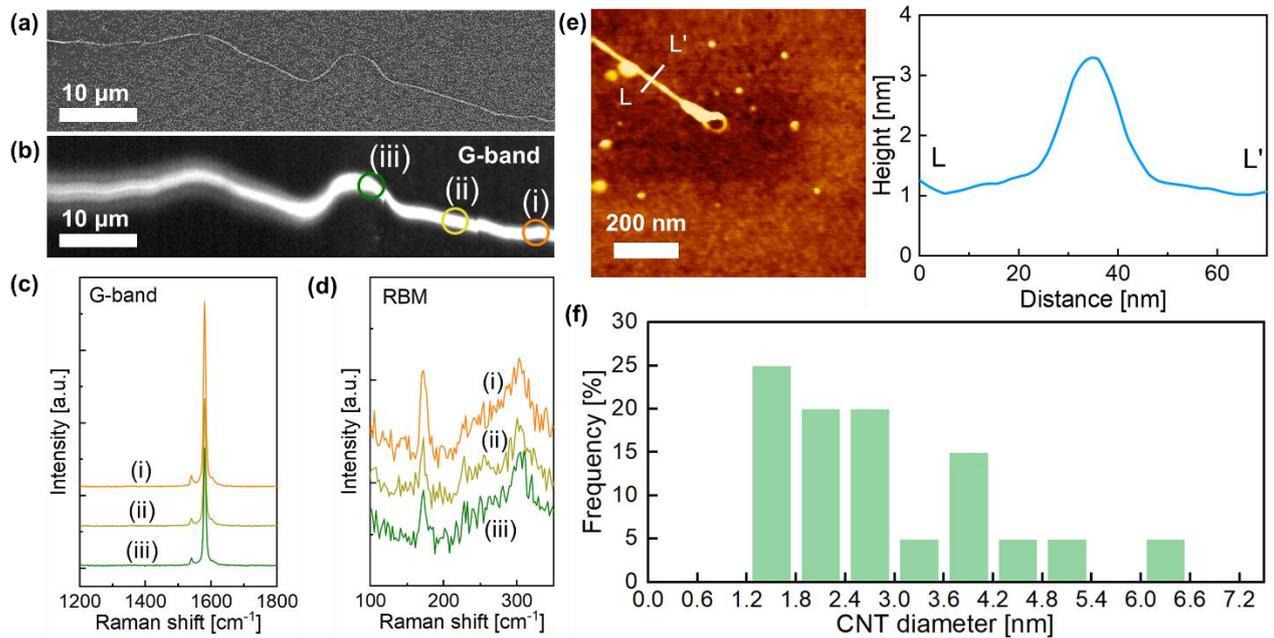

**Figure 6** (a) SEM and (b) G-band Raman mapping images of an aligned CNT. The excitation wavelength for the Raman measurement is 532 nm. Raman spectra of (c) G-band and (d) RBM obtained at the circles in (b) marked as (i), (ii), and (iii). (e) AFM image and line profile of the same CNT as (a) and (b) measured near its tip. The line profile was obtained along a segment of the AFM image represented by a line between two points, denoted as L and L′. (f) Diameter distribution of CNTs formed by the kite-growth mechanism from nonmetallic growth seeds.

## 4. Conclusion

Herein, we achieved kite growth of long, aligned CNTs from nonmetallic growth seeds by treating ND particles and controlling the conditions in the reactor. The temperature and flow rate dependence of the length and yield of aligned CNTs indicate the importance of the gas flow condition for initiating the kite growth from nonmetallic growth seeds, as in the case of metal catalysts. Using the SE yield analysis of CNT tips for the ND-deposited samples and the control Fe-deposited samples, we excluded the possibility of metal impurities affecting the aligned CNT growth from ND. A combined analysis of SE yield maps and AFM was used to detect the existence of NPs at the CNT tips, which confirms the tip-growth mode of CNTs grown from nonmetallic NPs. Although direct characterizations are expected to confirm their identity in the future, these nonmetallic NPs are presumed to be NDs or carbon NPs derived from NDs. Structural characterization revealed the high crystallinity of the CNTs and their relatively small diameters. This study is the first to demonstrate the use of nonmetallic growth seeds for kite growth and will provide alternative starting materials for fabricating long, aligned CNTs. As nonmetallic NPs have no catalytic effect, the yield of the grown CNTs lower than that from metal-catalyzed kite growth, and this also results in trace amounts of impurities that can still grow a significant percentage of CNTs. Thus, improving the efficiency of nonmetallic NPs in CNT growth is one of the important aspects of future research. With these challenges resolved, nonmetallic NPs will be more suitable for growing aligned CNT arrays due to the advantages of nonmetallic NPs over metal catalysts, including the high elemental purity with virtually no metal NPs in the resulting CNTs and the capability of growing CNTs on various substrates. We believe that the realization of kite-grown CNTs from nonmetallic NPs would assist in the practical applications of aligned CNT-based devices.

## Acknowledgments

The authors would like to thank Dr. T. Sakata of the Research Center for Ultra-High Voltage Electron Microscopy, Osaka University, for assistance in the SEM observation. A part of this work was financially supported by JSPS, KAKENHI (Grant Numbers JP15H05867 and JP17H02745). A part of this work was conducted at the Advanced Research Infrastructure for Materials and Nanotechnology Open Facilities, Osaka University (Grant Number JPMXP1222OS1008).

arrays: growth mechanism, controlled synthesis, characterization, properties and applications, Chem. Soc. Rev. 46 (2017) 3661–3715. https://doi.org/10.1039/C7CS00104E.

[13]  M. He, S. Zhang, J. Zhang, Horizontal Single-Walled Carbon Nanotube Arrays: Controlled Synthesis, Characterizations, and Applications, Chem. Rev. 120 (2020) 12592–12684. https://doi.org/10.1021/acs.chemrev.0c00395.

[14]  Y. Zhang, A. Chang, J. Cao, Q. Wang, W. Kim, Y. Li, N. Morris, E. Yenilmez, J. Kong, H. Dai, Electric-field-directed growth of aligned single-walled carbon nanotubes, Appl. Phys. Lett. 79 (2001) 3155–3157. https://doi.org/10.1063/1.1415412.

[15]  A. Ismach, L. Segev, E. Wachtel, E. Joselevich, Atomic-Step-Templated Formation of Single Wall Carbon Nanotube Patterns, Angewandte Chemie International Edition. 43 (2004) 6140–6143. https://doi.org/10.1002/anie.200460356.

[16]  H. Ago, K. Nakamura, K. Ikeda, N. Uehara, N. Ishigami, M. Tsuji, Aligned growth of isolated single-walled carbon nanotubes programmed by atomic arrangement of substrate surface, Chemical Physics Letters. 408 (2005) 433–438. https://doi.org/10.1016/j.cplett.2005.04.054.

[17]  R. Zhang, Y. Zhang, Q. Zhang, H. Xie, W. Qian, F. Wei, Growth of Half-Meter Long Carbon Nanotubes Based on Schulz–Flory Distribution, ACS Nano. 7 (2013) 6156–6161. https://doi.org/10.1021/nn401995z.

[18]  S. Huang, M. Woodson, R. Smalley, J. Liu, Growth Mechanism of Oriented Long Single Walled Carbon Nanotubes Using "Fast-Heating" Chemical Vapor Deposition Process, Nano Lett. 4 (2004) 1025–1028. https://doi.org/10.1021/nl049691d.

[19]  Z. Jin, H. Chu, J. Wang, J. Hong, W. Tan, Y. Li, Ultralow Feeding Gas Flow Guiding Growth of Large-Scale Horizontally Aligned Single-Walled Carbon Nanotube Arrays, Nano Lett. 7 (2007) 2073–2079. https://doi.org/10.1021/nl070980m.

[20]  T. Tsuji, K. Hata, D.N. Futaba, S. Sakurai, Additional obstacles in carbon nanotube growth by gas-flow directed chemical vapour deposition unveiled through improving growth density, Nanoscale Adv. 1 (2019) 4076–4081. https://doi.org/10.1039/C9NA00209J.

[21]  Z. Zhu, N. Wei, W. Cheng, B. Shen, S. Sun, J. Gao, Q. Wen, R. Zhang, J. Xu, Y. Wang, F. Wei, Rate-selected growth of ultrapure semiconducting carbon nanotube arrays, Nat Commun. 10 (2019) 4467. https://doi.org/10.1038/s41467-019-12519-5.22

Supporting information

# Gas flow–directed growth of aligned carbon nanotubes from nonmetallic seeds

*Yuanjia Liu [a], Taiki Inoue [a], Mengyue Wang [a], Michiharu Arifuku [b],*

*Noriko Kiyoyanagi [b], Yoshihiro Kobayashi [a]*

a Department of Applied Physics, Osaka University, Suita, Osaka 565-0871, Japan

b Nippon Kayaku Co., Ltd., Kita-ku, Tokyo 115-8588, Japan

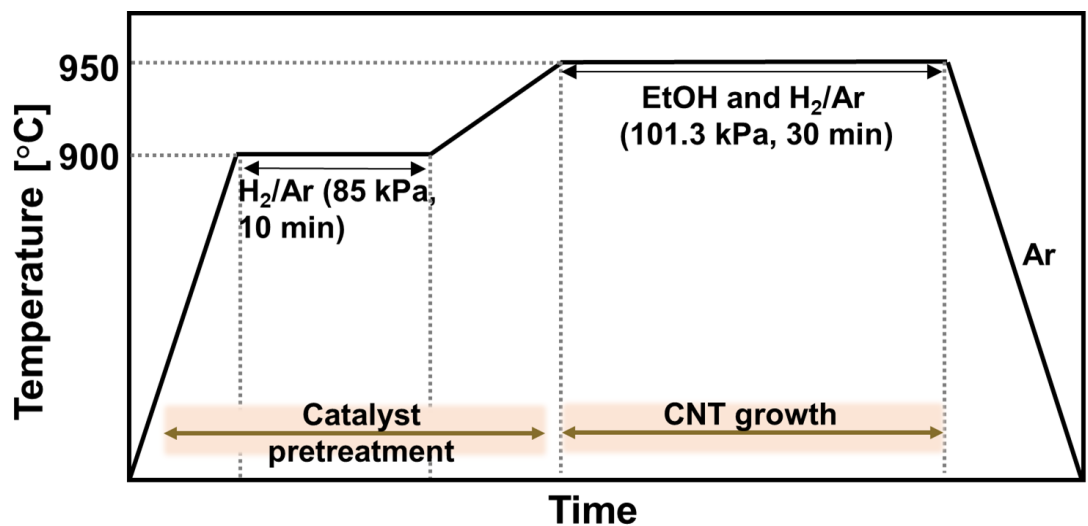

**Figure S1** Schematic of the kite-growth process of CNTs from a metal catalyst (Fe). Fe catalyst was patterned by a dip-pen method using $FeCl_3$ ethanol solution (0.03 mol/l) and a wooden stick. The CNT growth was initiated by introducing 70 sccm of $H_2$ (3%)/Ar bubbling through an ethanol reservoir at 40°C and 840 sccm of $H_2$ (3%)/Ar through another gas line.

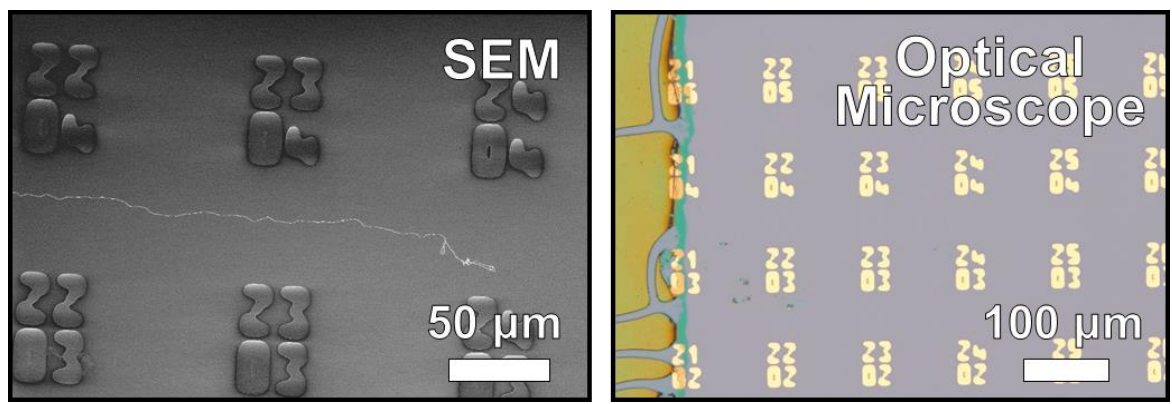

**Figure S2** SEM and optical microscopy images of the same region where a target CNT exists. The position was adjusted by number-shaped metal markers on the substrate. The vertical boundary toward the left in the optical microscopy image indicates the ND-deposited area (left) and the aligned CNT growth area without ND (right).

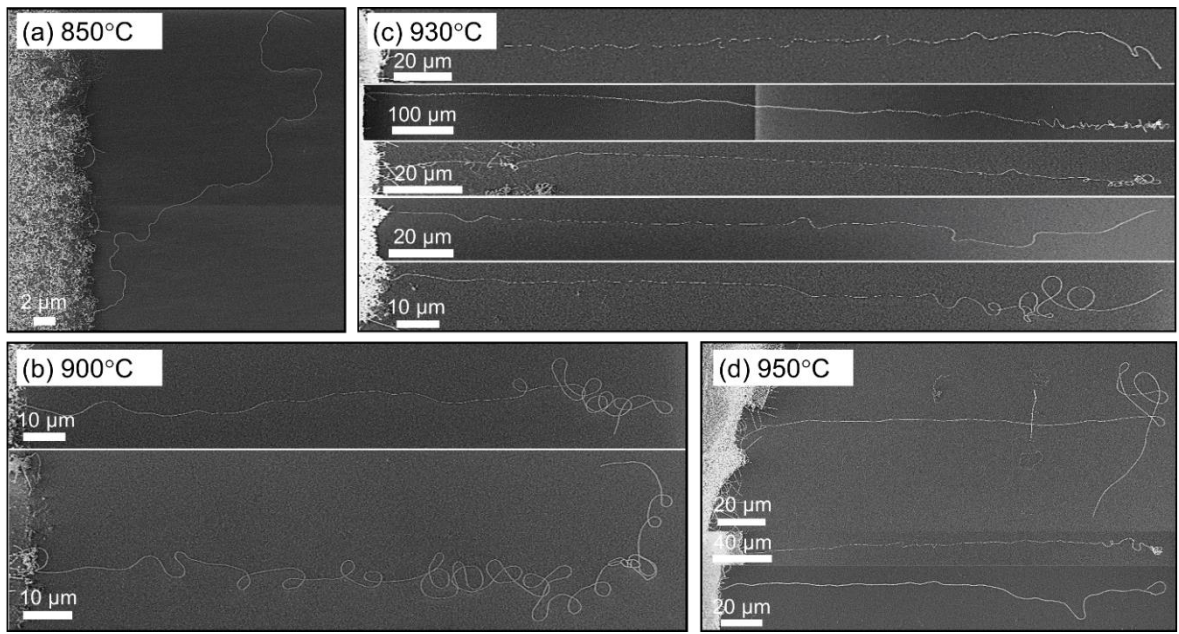

**Figure S3** SEM images of the aligned CNTs grown on the SiO$_2$/Si substrate at different temperatures with a flow rate of 455 sccm: (a) 850°C, (b) 900°C, (c) 930°C, and (d) 950°C.

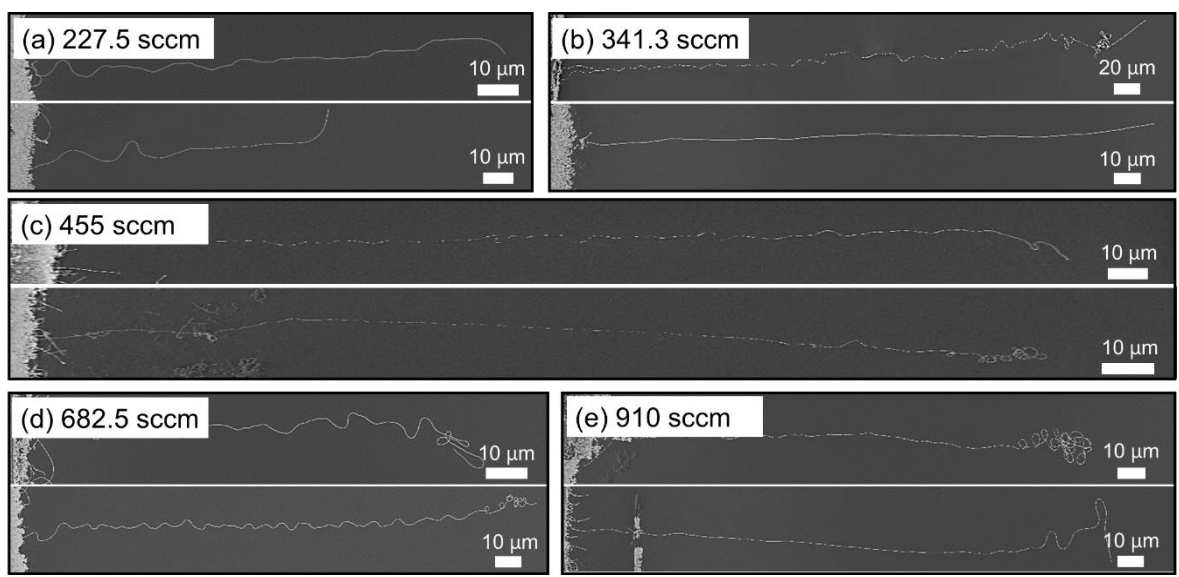

**Figure S4** SEM images of the aligned CNTs grown on the SiO$_2$/Si substrate with different flow rates at 930°C: (a) 227.5 sccm, (b) 341.3 sccm, (c) 455 sccm, (d) 682.5 sccm, and (e) 910 sccm.

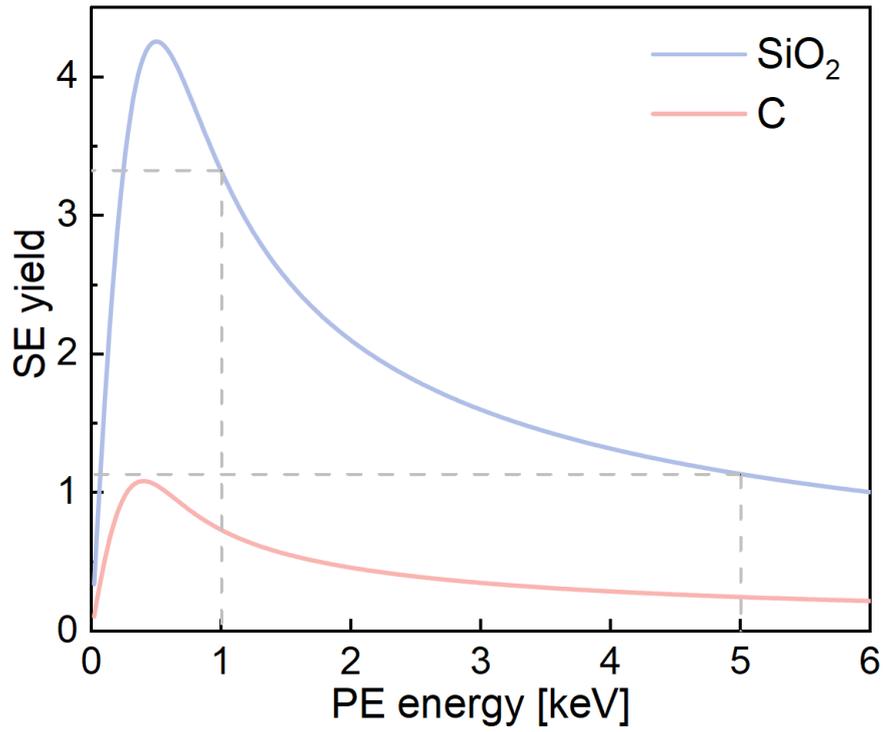

**Figure S5** Plot of SE yield at different PE energies. The calculation formula and data for C [1] and the data for SiO$_2$ [2] were obtained from previous studies.

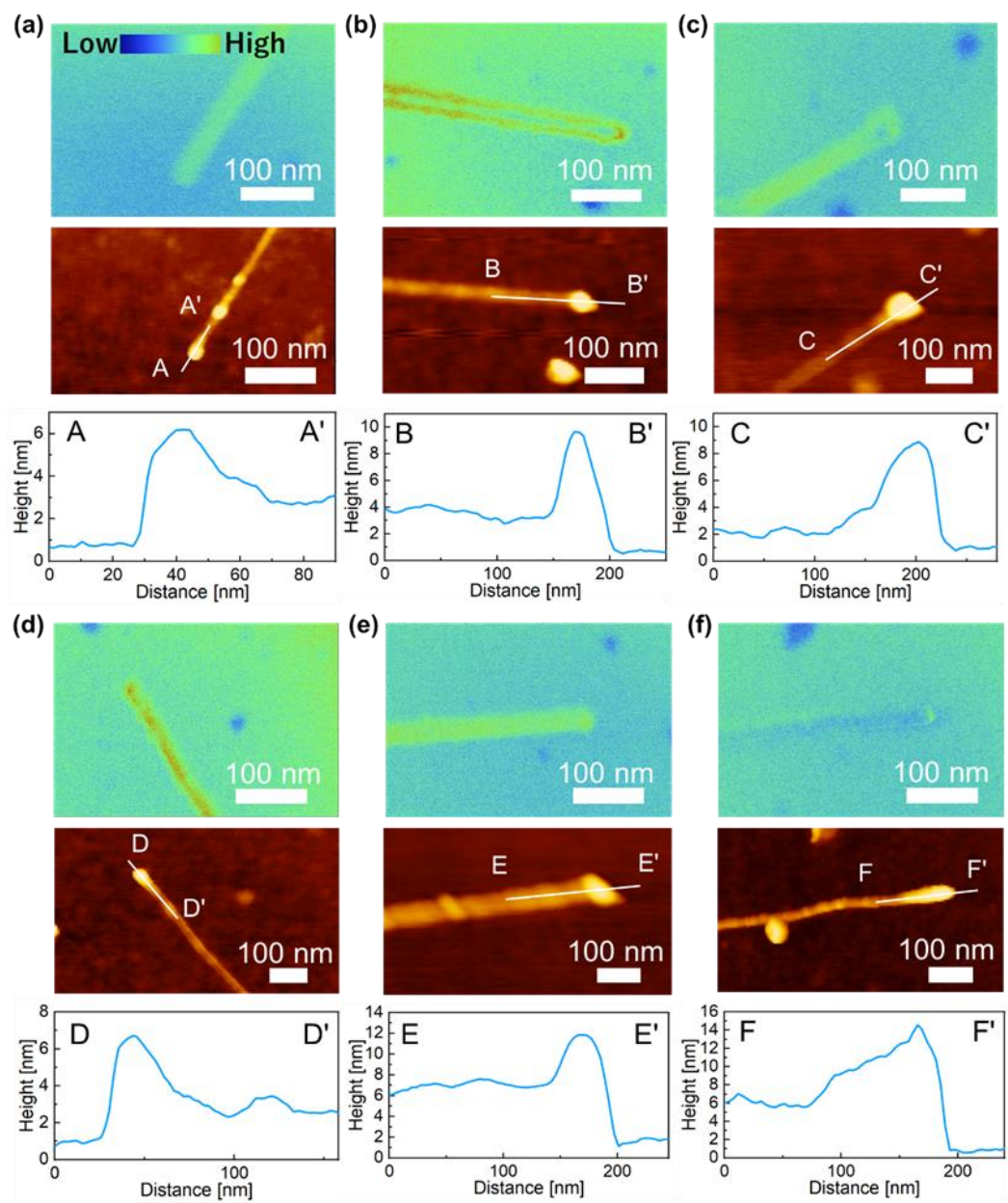

**Figure S6** SE yield maps and AFM images and line profiles of aligned CNT tips (>50 μm). The line profile was obtained along a segment of the AFM image represented by a line between two points. While the presence of metal nanoparticles was denied by SE yield maps, the presence of some kind of nanoparticles at the tip of CNTs was confirmed by AFM, suggesting that CNTs were grown from nonmetallic NPs via the tip-growth mode. Note that some particles surrounding the CNTs can be attributed to the formation of amorphous carbon during the CVD process.

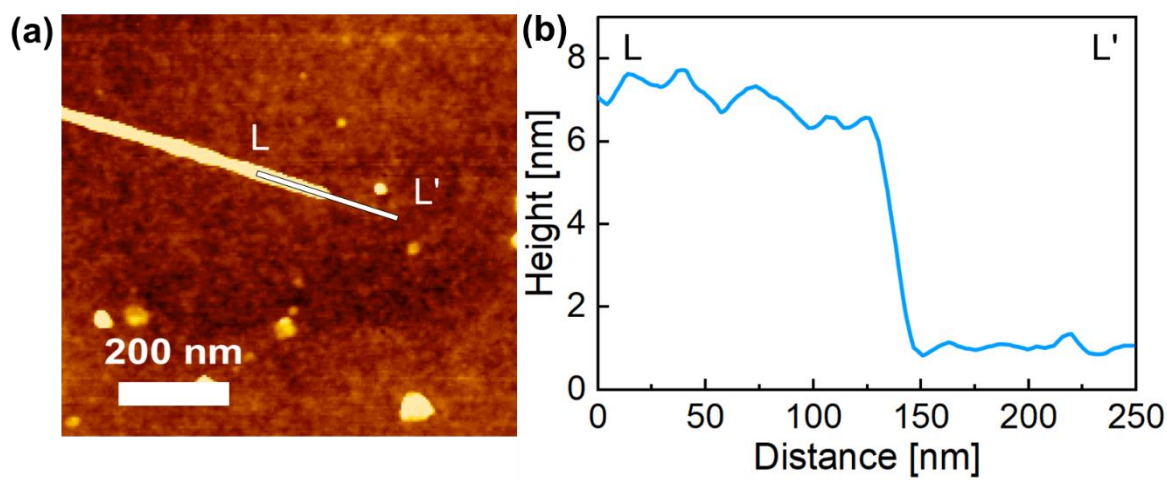

**Figure S7** (a) AFM image and (b) line profile of the CNT. The line profile was obtained along a segment of the AFM image represented by a line between two points, denoted as L and L′.

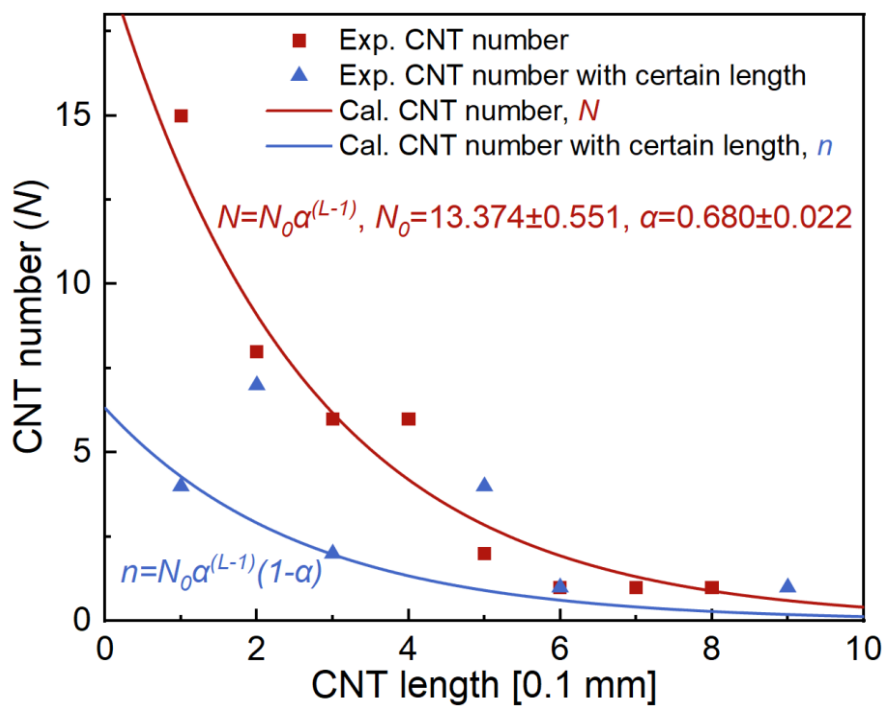

**Figure S8** The number distribution of CNTs grown from nonmetallic NPs via kite growth.

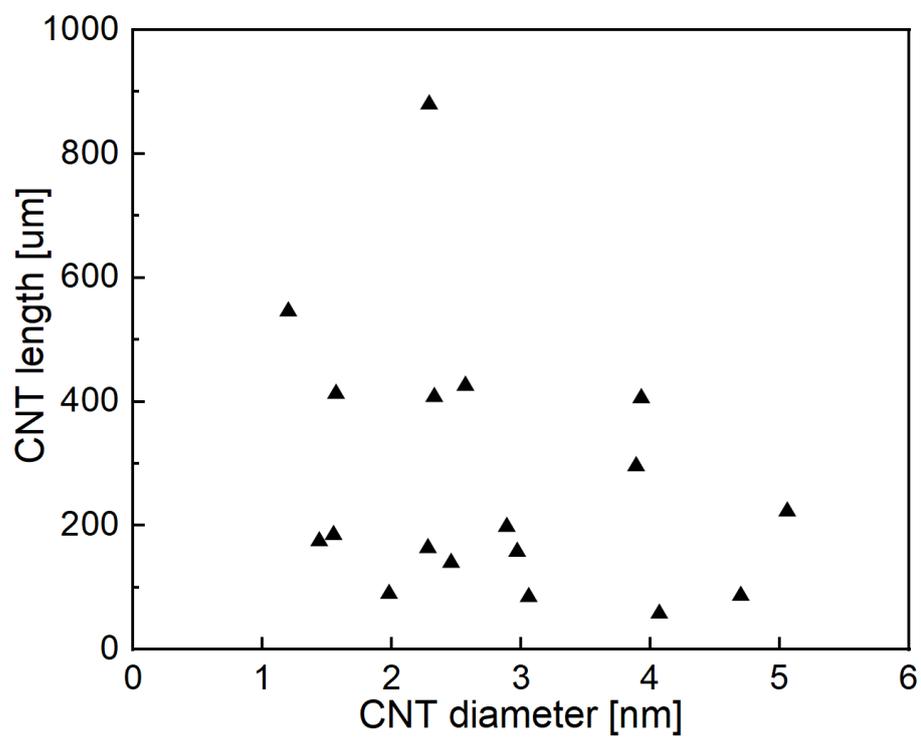

**Figure S9** The distribution of CNT diameter and the corresponding CNT length grown from nonmetallic NPs via kite growth.